\documentclass[11pt]{article}

\usepackage[margin=1in]{geometry}
\usepackage{amsmath}
\usepackage{amssymb}
\usepackage{graphicx}
\usepackage{booktabs}
\usepackage{setspace}
\usepackage{xcolor}
\usepackage[hidelinks]{hyperref}
\usepackage[utf8]{inputenc}
\usepackage{textcomp}
\usepackage[section,below]{placeins}
\title{\textbf{A robust association between LLM use and scientific productivity: Assessing stopping-time selection}}
\author{
Response to Renault, Bergeaud, and Bosquet\\
{Keigo Kusumegi, Xinyu Yang, Paul Ginsparg, Mathijs de Vaan, Toby Stuart, Yian Yin}}
\date{\today}

\begin{document}
\maketitle

\begin{abstract}
\noindent Renault, Bergeaud, and Bosquet (hereafter RBB) argue that dating LLM
adoption as the first month in which an author's abstract is flagged induces a stopping-time selection that can produce a positive event-study path even when there is no causal effect~\cite{renault2026comment}. Although this mechanism is mathematically possible, it does not constitute proof of a null effect. Recalibrating RBB's own random placebo to the detector's realized flag rate, we show that the measured association stays well above this benchmark, so the artifact is too small to explain the productivity changes. We further re-estimate the association between LLM adoption and productivity with a series of complementary designs in which the timing artifact cannot bias the estimate: a before-and-after comparison that dates adoption in one year and measures output in another, a conservative control group for difference-in-differences, an intensity-based specification that never defines an adoption date, and a rank-based measurement holding the flag rate fixed. A positive productivity association persists across all of these estimates, while the same tests run on pre-ChatGPT placebo data return null effects. The artifact RBB identify is real but bounded, and it does not account for the pattern we report.
\end{abstract}

\noindent We thank RBB for their careful engagement with our work~\cite{renault2026comment}. Their comment raises an important question about event-study designs that date treatment from the outcome, which we address directly in this reply. Before turning to the technical details, we situate RBB's comment within our original paper~\cite{kusumegi2025scientific}. Our paper reports several distinct findings; RBB's comment bears on one alone---the event-study estimate of an LLM--productivity association, obtained from an adopter/non-adopter comparison of the kind common in existing literature~\cite{hao2026artificial,filimonovic2025can} ---and does not concern the paper's other findings.
We also highlight that the original paper was explicit about the limitations of dating adoption from first detected use and about the well-known difficulty of separating adoption from output, and the paper cautioned against a causal reading of the effect magnitude on those grounds. RBB's thoughtful comment extends these cautions.

RBB argue that, when LLM adoption is dated to the first month in which an author's abstract is flagged, the month before adoption is, by construction, a month with no detected flag, which makes it an unusually low-output month (discounted by a factor of $1-p$). Comparing the months that follow against this depressed reference can produce a positive post-treatment path, even in the absence of a real productivity effect. Although the mechanism RBB identify is plausible, it establishes only that such an artifact can arise. How much of the estimated association it actually explains is an empirical question, and it is the one we take up here.

We conducted several analyses to address this question. First, we show that the placebo test proposed by RBB is biased toward the ``real'' effect, so an apparently similar shape does not constitute evidence for a null effect. Second, since the size of the artifact is governed by how often the detector flags a paper, we recalibrate the random placebo's ``firing rate'' to match the LLM detector's; even against this inflated baseline, the measured association remains substantially larger. Third, we re-estimate the association under four designs in which first-detection dating cannot bias the estimate: a positive association persists across all four, while the same procedures applied to pre-ChatGPT placebo data return null effects. We describe each point in detail below.

\section{A matching placebo does not establish a null}
RBB write that, because their placebo flags are uninformative, any post-treatment pattern that arises from them must reflect the timing rule alone. Two features of their placebos go against this claim. Both inflate the placebo path above the underlying artifact, so the placebo overstates rather than measures the artifact.

First, reproducing the shape of the event-study does not, by itself, establish a null. If LLM adoption elevates scientific production from a low baseline, almost any paper-level marker, real or random, would locate the adoption month from the surge in papers. The placebo's assignment of treatment status and timing would then largely coincide with the real one, and the placebo would trace the real path even when the true empirical association is large -- not because the flag is informative, but because under a large effect the surge makes any output-correlated flag a noisy copy of the true assignment. A placebo that matches the baseline shape is therefore consistent with a real effect and cannot by itself demonstrate its absence.

\begin{figure}[!h]
\centering
\includegraphics[width=\textwidth]{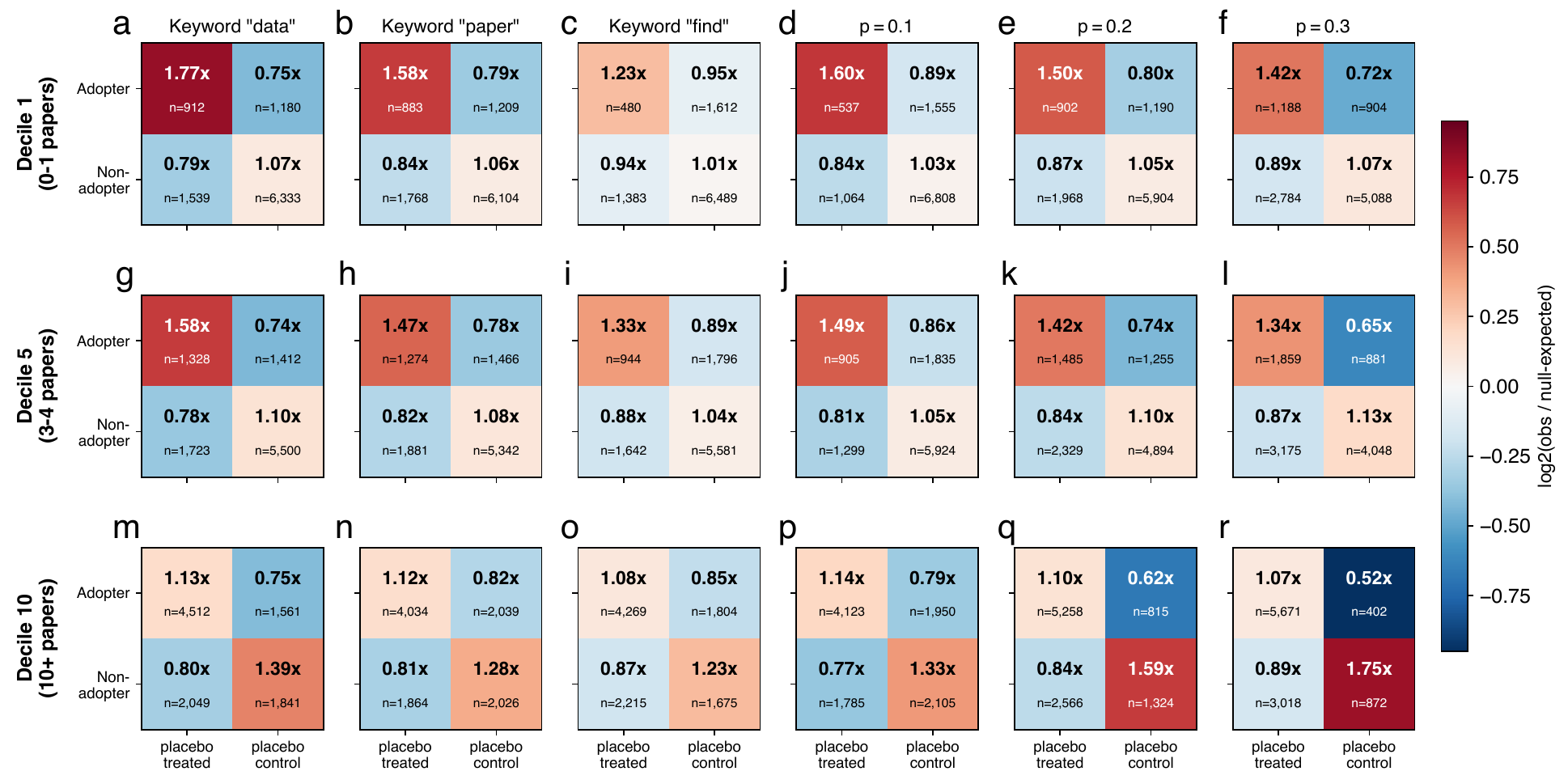}
\caption{\textbf{Adopters are over-represented among the placebo-treated across different levels of baseline output.} For each of six placebo rules (three keyword flags, three random flags at $p=0.1,0.2,0.3$), authors are cross-tabulated by actual adopter status against placebo-treated status, and each cell is compared to an independence-shuffle null that permutes the placebo label while holding both marginals fixed. Authors are first split based on \emph{pre-ChatGPT} (2020--2021) publication volume, and the cross-tab and null are recomputed \emph{within} each subgroup. Because both the real detector and the placebo flags increase with output, some overlap is expected from historically prolific authors triggering both. But the over-representation does not collapse once baseline output is held fixed: it
persists across subgroups and placebo rules (the color shrinkage at the top reflects marginal saturation, not a weaker association).}
\label{fig:confusion}
\end{figure}

Second, we show this contamination is present in our data. Because both the detector and the placebo flags are increasing in output, high-output authors are over-selected into every placebo: authors classified as LLM adopters are over-represented among the placebo-treated by a factor of roughly 1.3 to 1.5 in aggregate, and correspondingly under-represented among the placebo-control; the enrichment persists within every subgroup of baseline output (Fig.~\ref{fig:confusion}).
The placebo does not isolate the pure stopping-time artifact. Because the placebo treatment is correlated with empirical adopter status even conditional on baseline output, its post-treatment path may also absorb whatever post-ChatGPT productivity association is carried by the empirical adopter classification.

This applies to the rate-matched random flag of Section~2 below: readers should keep in mind it is contaminated by the same output-driven selection and therefore also sits above the pure artifact, which makes the comparison conservative. 

\section{A conservative placebo benchmark still leaves a positive excess}
RBB establish that the magnitude of the artifact is a function of the flag rate (i.e. the share of papers the detector marks as LLM-assisted). They derive a closed form in which the post-treatment plateau equals $-\log(1-p)$, a function of the flag rate alone, and they confirm it by running random flags at increasing rates and showing that the estimate rises with the rate. We build on this directly. Because the size of the artifact is set by the flag rate, a comparison between the detector and an uninformative flag is meaningful only when both are set at the same rate. RBB's placebos are defined at rates that deviate from the detector's, so a raw comparison of magnitudes confounds the flag rate with any real signal. 

We instead hold the rate fixed at the detector's own realized rate: like RBB, we run a random flag that carries no information about LLM use, but we calibrate it to the rate of the detector and apply the identical first-detection event study to it. The path it traces provides a conservative benchmark for the timing artifact, calibrated directly to the detector's actual flagging frequency.

\begin{figure}[!hbp]
\centering
\includegraphics[width=\textwidth]{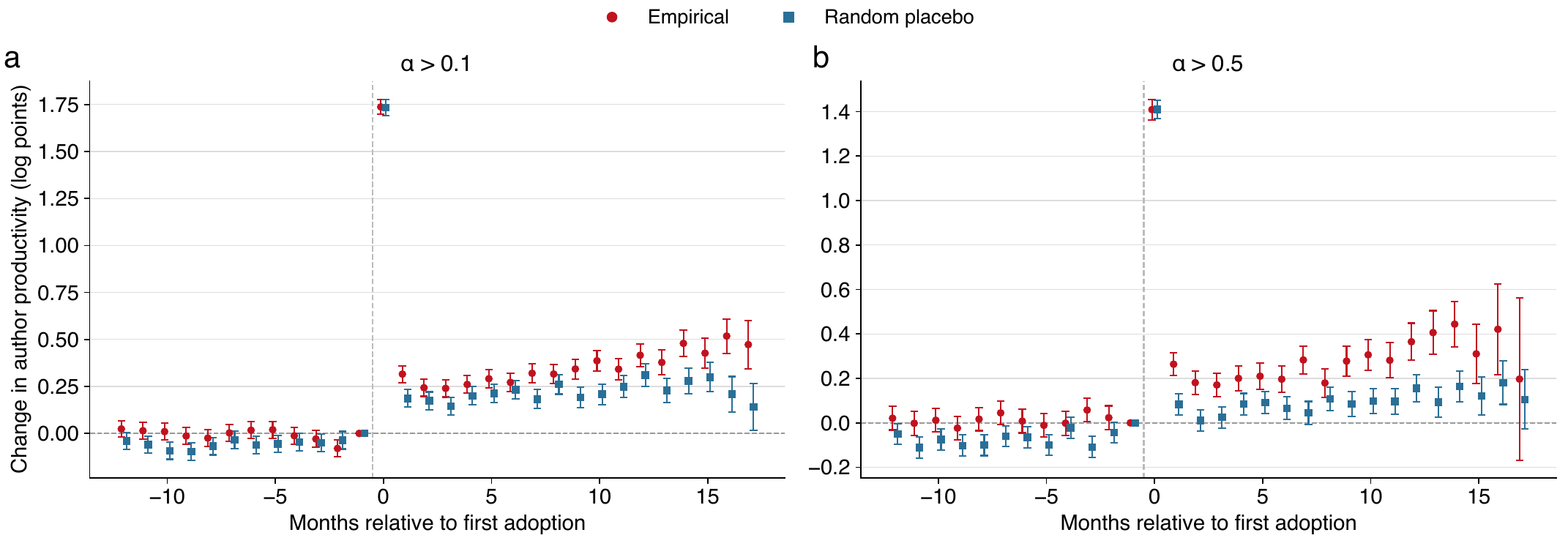}
\caption{\textbf{The LLM detector exceeds the placebo matched on the realized flag rate.}
Event-study coefficients for the empirical LLM detector and for a random, uninformative flag calibrated to the rate of the detector, at the $\alpha>0.1$ and $\alpha>0.5$ classification thresholds. Holding the flag rate fixed isolates the comparison of interest: the treatment-month ($k=0$) coefficients coincide, confirming that the artifact is held fixed, and the detector's path nonetheless sits above the placebo's throughout the post-treatment period, with the gap widening at the stricter threshold.}
\label{fig:ratematch}
\end{figure}

When the real LLM detector and the random placebo are set to flag papers at the same rate, their treatment-month ($k=0$) coefficients align closely in practice, providing a visual check that the baseline selection environment is properly matched. After matching, the detector's path sits above the benchmark across the entire post-treatment period (Fig.~\ref{fig:ratematch}). The average post-treatment excess of the detector over the matched benchmark is 0.119 (95\% CI $[0.068, 0.169]$) at the $\alpha>0.1$ threshold (realized rate $p^{\ast}=0.147$) and 0.183 (95\% CI $[0.121, 0.244]$) at the stricter $\alpha>0.5$ threshold ($p^{\ast}=0.054$). Thus, the excess is largest at the stricter threshold, where the detector most cleanly picks up LLM use. Importantly, because this calibrated placebo is a conservative benchmark for the narrow stopping-time mechanism (as detailed in Section 1), the resulting excess provides a conservative estimate relative to this benchmark. 

\begin{figure}[!h]
  \centering
  \includegraphics[width=\linewidth]{
  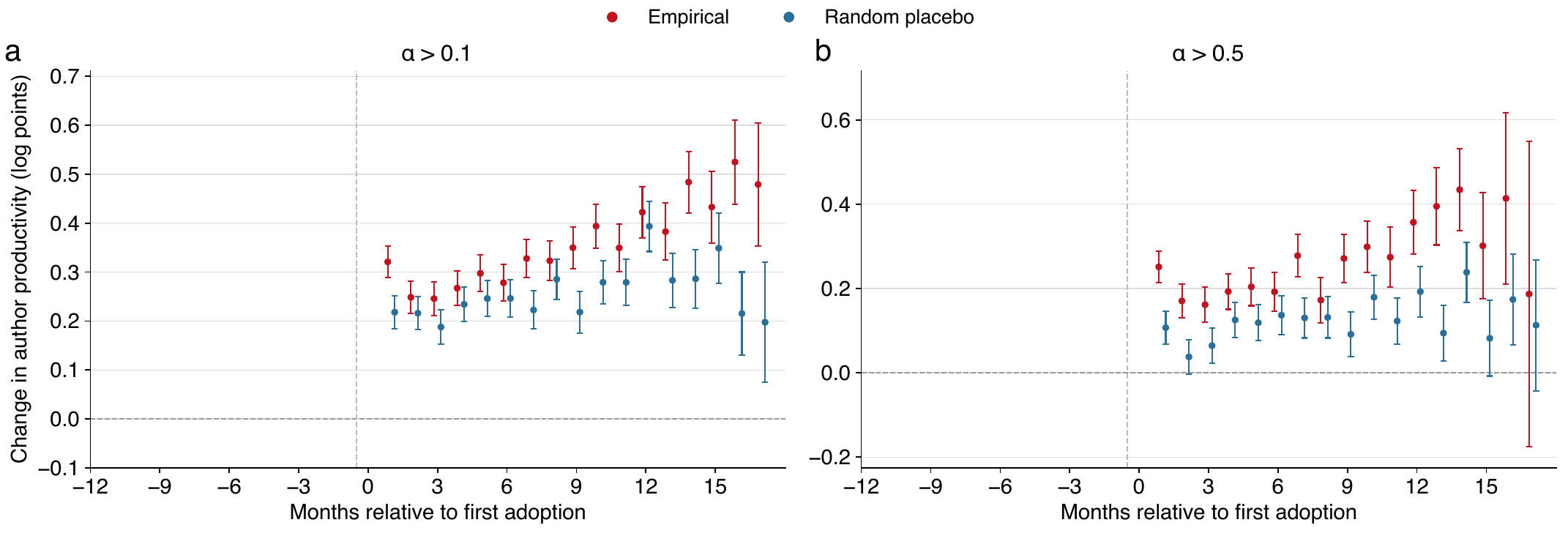}
  \caption{\textbf{The LLM detector exceeds the placebo matched on the first-detection spike, using the pooled pre-treatment periods as the baseline.}
Event-study coefficients from the controlled stacked difference-in-differences for the empirical LLM detector (red) and for a random, uninformative Bernoulli flag (blue), at the $\alpha>0.1$ and $\alpha>0.5$ classification thresholds. The random flag's rate is calibrated so that its treatment-month ($k=0$) first-detection spike equals the detector's, rather than to its overall flag rate. The entire pre-period ($k\le-1$) is pooled into the estimator's reference category, so both arms share an identical baseline ($\equiv 0$ by construction); the matched $k=0$ spikes (${\approx}1.7$ and ${\approx}1.4$ log points) are off-scale and omitted from the plotted range. Holding both the pooled baseline and the first-detection spike fixed isolates the comparison of interest: any divergence at $k\ge1$ is the residual that the spike-matched mechanical null cannot reproduce. The detector's path nonetheless sits above the placebo's throughout the post-treatment period, with the gap widening at the stricter threshold ($+0.10$ log points at $\alpha>0.1$, $+0.13$ at $\alpha>0.5$).}
  \label{fig:prepool}
\end{figure}

Note that the random placebo group exhibits negative coefficients for most pre-treatment periods, which is absent in our real LLM flag. This is consistent with the idea that the seemingly similar shape does not demonstrate a null productivity effect. To that end, we consider a similar exercise using all pre-treatment periods (instead of the $k=-1$ period) as the baseline, recalibrating our $p^{\ast}$ and again finding similar patterns (Fig.~\ref{fig:prepool}).

\FloatBarrier
\section{Additional evidence from four alternative designs}
 
The artifact arises when post-detection output is compared against a reference month that first detection has mechanically depressed. We therefore re-estimate the association under four designs in which dating adoption from the first detected month cannot mechanically bias the estimate: the depressed reference either does not enter the estimand 
(Sections~\ref{ss:beforeafter} and~\ref{ss:intensity}), or the same depression arises in the comparison arm and differences out (Sections~\ref{ss:control} and~\ref{ss:rank}). The specific stopping-time mechanism identified by RBB therefore cannot generate the reported contrast in any of these designs. Each design still returns a positive association in the data and nothing on pre-ChatGPT placebo data. These estimates are largely consistent, and should not be interpreted causally. Their common purpose is to test whether the positive association survives when the specific depressed-reference mechanism is removed or incorporated into the comparison.
 
\subsection{A before-and-after comparison across separated periods}\label{ss:beforeafter}
\begin{figure}[!h]
\centering
\includegraphics[width=0.75\textwidth]{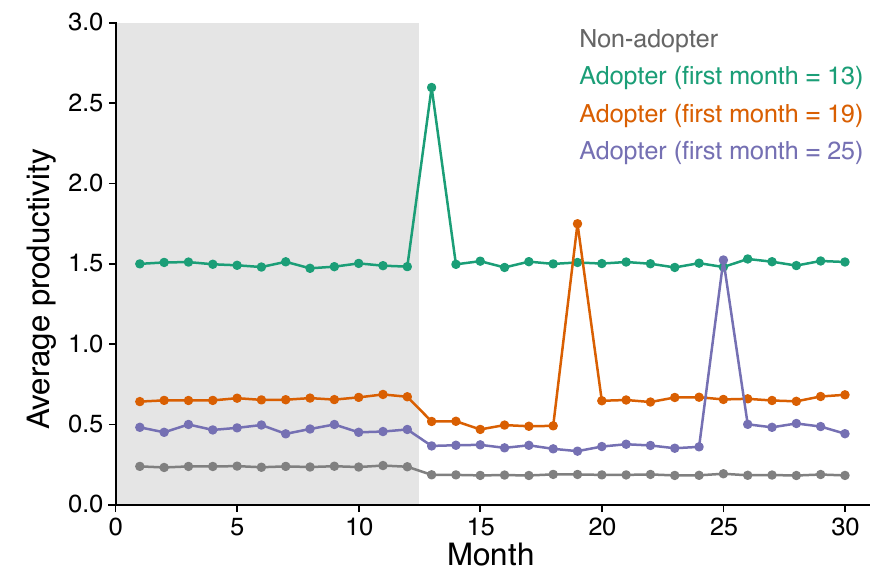}
\caption{\textbf{Productivity dynamics under the simulated series.}
We replicated the simulation exercise in RBB's Fig.~2. Despite the stopping-time event, comparing adopters' post-adoption productivity versus their productivity in the pre-LLM era recovers an unbiased estimate of the treatment effect (in this case, returning approximately zero under the simulated null).}
\label{fig:simulation}
\end{figure}

We build on RBB's simulation (their Fig.~2) and decompose their difference-in-differences estimate by adoption cohort. Fig.~\ref{fig:simulation} plots average productivity for three adopter cohorts and for never-adopters under a simulated null. The stopping-time artifact is visible as a single-month spike at each cohort's adoption month---but productivity returns immediately to its pre-adoption level, with no sustained shift. Comparing a post-adoption window against the pre-LLM era therefore returns approximately zero when the true effect is zero: the artifact does not survive a comparison across separated periods. 

We then apply the identical comparison to the data. We classify authors as adopters on the basis of their 2023 detector flags and compare their average monthly output in January--June 2024 against January--June 2022. Output rises by about 17.3\% at the $\alpha>0.1$ threshold and 25.1\% at the $\alpha>0.5$ threshold. The same exercise on the pre-ChatGPT placebo returns approximately $-0.9\%$ and $0.0\%$, respectively.
Because the design returns nothing under the simulated null and nothing on the pre-ChatGPT placebo, the positive estimate it yields in the data cannot be attributed to the stopping-time artifact.

This before-and-after comparison removes the stopping-time artifact, since adoption is dated in one period and output measured in another, but it does so by giving up a contemporaneous control group, and is therefore not immune to system-level temporal changes between 2022 and 2024. Any secular shift in submission rates over this window may be absorbed into the estimate, since the design holds the cohort fixed rather than calendar time. We present this comparison only to show that the artifact does not survive a design that separates the dating period from the measurement period, and the result should be interpreted as corroborative.

\FloatBarrier
\subsection{A conservative control group}\label{ss:control}

A second approach retains the event-study timing but revises the control group. The artifact is an asymmetry: dating adoption from the first detected month makes the treated author's reference period a low-output month (discounted by $1-p$), while post-adoption months return to baseline (multiplied by the LLM boosting effect, if any). Never-treated authors are discounted by $1-p$ every month, so their flat profile cannot offset the treated jump and the artifact survives differencing. Combining never-treated with not-yet-treated authors fixes this: the combined population carries a similar selection, so the pooled control group's reference period is discounted by $1-p$ while its later months are not---the same asymmetry as the treated. Adding them in reproduces the mechanical jump in the comparison arm, allowing it to difference out.

This control group is conservative by construction. Not-yet-treated authors are
future adopters, so if LLM use raises output, they will exhibit the same boost after 
adoption, which pulls the control mean upward and attenuates the DiD estimate toward zero.
Under a true null the specification returns zero; under a true effect it
underestimates rather than inflates the association. A significant positive difference is therefore a lower bound on the association, not an artifact of the design.

\begin{figure}[!h]
\centering
\includegraphics[width=\textwidth]{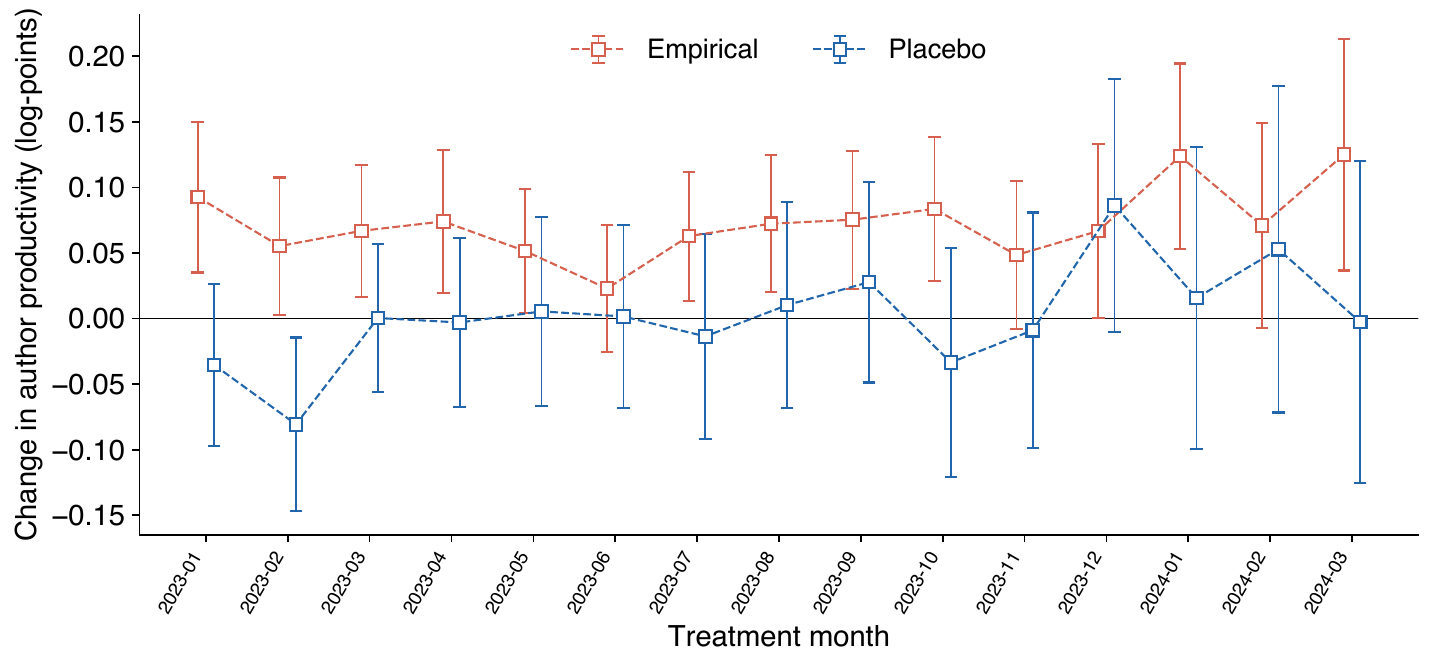}
\caption{\textbf{A conservative control group leaves a positive effect that the
  placebo does not reproduce.} Monthly $2\times2$ difference-in-differences
  coefficients (log-points) for treated authors matched to never-treated and
  not-yet-treated controls on treatment-month productivity, across treatment
  months $t \in [13, 27]$ (red), with the identical estimator on the pre-ChatGPT
  placebo (blue). The empirical coefficient is positive throughout and significant
  in the majority of months; the placebo coefficient is centered near zero and
  individually insignificant in every month. Because not-yet-treated controls are
  future adopters, this specification attenuates the estimate toward zero, so the
  surviving positive coefficient is a lower bound.}
\label{fig:conservative}
\end{figure}

For each treatment month $t \in [13, 27]$, we match the treated group---authors
with their first detected LLM-assisted paper in month $t$---to a control group of
authors with the same month-$t$ productivity and no detected LLM-assisted papers
up to $t$. Figure~\ref{fig:conservative} reports the resulting
$2\times2$ difference-in-differences coefficient for each month, alongside the
identical estimator applied to the pre-ChatGPT placebo. The empirical coefficient is positive across the entire treatment window, concentrated around $0.05$ to $0.13$ log-points, with confidence intervals that exclude zero in most months (pooled estimate: $0.068$ log-points, 95\% CI $[0.053, 0.083]$). The placebo coefficient, by contrast, is individually insignificant in every month---its confidence interval includes zero throughout. Because this control group attenuates the estimate rather than inflating it, the persistent positive coefficient is a lower bound on the association and cannot be the stopping-time artifact.

\FloatBarrier
\subsection{An intensity specification with no adoption date}\label{ss:intensity}
Following recent literature~\cite{daniotti2026using}, we estimate an alternative specification: regress productivity in the current quarter $y_{u,q}$ on the intensity of an author's recent AI use, $AI_{u,q-1}$, 
measured as the average detector score, the share of papers with $\alpha>0.1$, and the share with $\alpha>0.5$ across the author's papers in the previous quarter:
$$y_{u,q}=\beta_{AI} AI_{u,q-1}+\rho_u+\tau_q+\epsilon_{u,q}$$

The lag breaks the mechanical link between contemporaneous flagging and contemporaneous output, and the specification does not define a discrete adoption month. As Table~\ref{tab:intensity} reports, every intensity measure is positive and highly significant in the data, and every measure is negative and insignificant in the pre-ChatGPT placebo (2020--2022 period). Because this specification traces within-author variation across periods, its coefficients are not directly comparable in scale to the difference-in-differences estimates, but its sign and significance corroborate the results from other designs.

\begin{table}[!h]
\centering
\caption{\textbf{Productivity increases with prior-period AI-use intensity, with no such pattern on pre-ChatGPT placebo data.} Each row regresses current productivity on a one-quarter-lagged measure of AI-use intensity. Coefficients are positive and significant in the data and negative and insignificant on the placebo.}\bigskip
\label{tab:intensity}

\begin{tabular}{lcc}
\toprule
Measure (one-quarter lag) & Empirical & Placebo \\
\midrule
Average detector score        & $+0.103$ ($p<0.001$) & $-0.086$ ($p=0.14$) \\
Share of papers $\alpha>0.1$   & $+0.031$ ($p<0.001$) & $-0.023$ ($p=0.17$) \\
Share of papers $\alpha>0.5$   & $+0.071$ ($p<0.001$) & $-0.101$ ($p=0.15$) \\
\bottomrule
\end{tabular}
\end{table}

\subsection{Rank-based measurement, holding the flag rate fixed}\label{ss:rank}
RBB's account implies that at a fixed flagging rate $p$, the $\alpha$ content of the flag is irrelevant: any rule firing at rate $p$ inherits the same timing artifact. A random $\mathrm{Bernoulli}(p)$ flag realizes exactly this artifact---it marks $(100p)\%$ of papers with no information about $\alpha$---and so serves as the artifact baseline at each rate. We hold $p$ fixed and compare two informative rules against it: flagging the top-$(100p)\%$ of papers by $\alpha$, and the bottom-$(100p)\%$. If $\alpha$ carries no signal, both should equal the random arm at every $p$.

\begin{figure}[t]
  \centering
  \includegraphics[width=\linewidth]{
  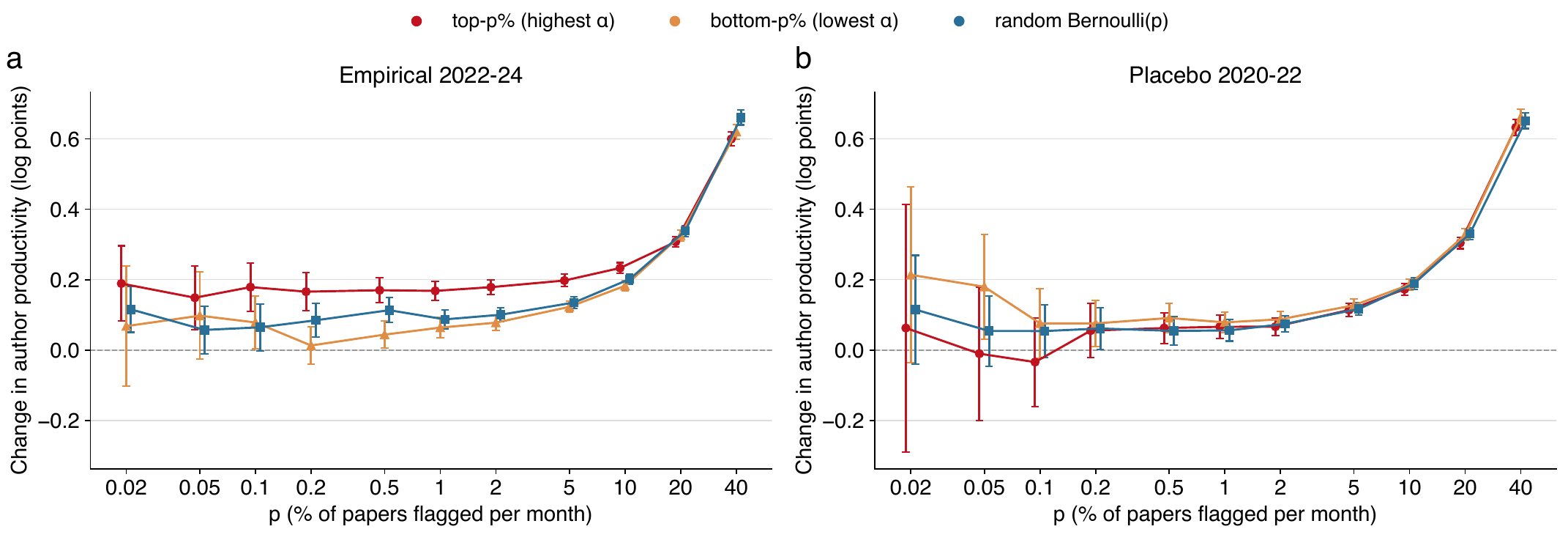}
  \caption{\textbf{At a fixed flagging rate, informative rules straddle the
  artifact baseline.} Each rule flags the same fraction $p$ of papers per month,
  so the stopping-time artifact is held fixed; the rules differ only in
  \emph{which} papers they mark---highest-$\alpha$ (top), lowest-$\alpha$
  (bottom), or a random $\mathrm{Bernoulli}(p)$ draw (the artifact baseline).}
  \label{fig:rank-based}
\end{figure}

Our results strongly reject this null (Fig.~\ref{fig:rank-based}). At low flagging rates, where the flagged set isolates the clearest cases of LLM assistance, the top-$\alpha$ rule sits above the random baseline and the bottom-$\alpha$ rule sits below it, differing by $15.3\%$ (at flagging rate $p = 0.2\%$). As $p$ rises the three rules tend to converge, since flagging more papers makes the selection rules coincide, and the $\alpha$ signal vanishes precisely where the rules stop being selective.

Because all three rules flag the same fraction of papers, the separation cannot be explained by the $p$-driven mechanical benchmark derived by RBB. Note that this estimate is again likely conservative: Highly productive authors will
occasionally produce a low-$\alpha$ paper and are thereby drawn into
the low-$\alpha$ condition, attenuating the contrast between the two arms. The
same exercise on the pre-ChatGPT placebo, where $\alpha$ carries no information
about LLM use, returns null-to-negative estimates in both conditions. The straddle pattern is thus specific to the period in which LLM
assistance is present.

\section{Discussion}

Across every design in which the stopping-time artifact cannot bias the estimate, a positive association between LLM use and productivity persists, while each corresponding pre-ChatGPT exercise returns null. An independent analysis using different data and identification reports the same sign~\cite{gartenberg2026more}. An artifact-based explanation of our result would therefore have to explain away not only the convergence of our own designs but also external evidence it does not touch. We read the agreement across unrelated approaches as the relevant signal.

We find the $p$-driven stopping-time component identified by RBB is real and quantitatively bounded, and it does not account for the positive association we report.
Benchmarked directly against an uninformative flag operating at the detector's own realized rate, the measured association remains above the artifact at both classification thresholds. We take the convergence of designs that would fail in different ways, were any of them artifact-driven, as evidence against a mechanical explanation.

We are explicit about what these results do and do not establish. The designs above address the stopping-time artifact, but they do not in any way resolve the endogeneity of technology adoption. As we acknowledge in the original article, who adopts, and when, is not random. As in our original paper, we do not claim that any estimate herein is causal. Rather, we establish that the specific artifact RBB describe is bounded, small relative to the measured association, and does not survive as an explanation once the comparison is made fair. The stopping-time mechanism is a useful caution for event-study designs that date treatment from the outcome. It does not support the conclusion that the association is an artifact.

A final point concerns what is being estimated. The object of our claim is the existence and direction of a positive association between LLM use and scientific output, not a specific magnitude. That magnitude is local by nature. It reflects a particular generation of models over a particular window of time, and it will change as the technology does. The question this exchange turns on is therefore whether a positive association survives once the dating artifact is accounted for, which the designs above establish, and not the precise size of any single coefficient.

\bibliographystyle{unsrt}
\bibliography{references}

@article{renault2026comment,
  title={Comment on Scientific production in the era of large language models},
  author={Renault, Thomas and Bergeaud, Antonin and Bosquet, Cl{\'e}ment},
  journal={arXiv preprint arXiv:2605.17979},
  year={2026}
}

@article{filimonovic2025can,
  title={Can GenAI improve academic performance? Evidence from the social and behavioral sciences},
  author={Filimonovic, Dragan and Rutzer, Christian and Wunsch, Conny},
  journal={arXiv preprint arXiv:2510.02408},
  year={2025}
}

@article{kusumegi2025scientific,
  title={Scientific production in the era of large language models},
  author={Kusumegi, Keigo and Yang, Xinyu and Ginsparg, Paul and de Vaan, Mathijs and Stuart, Toby and Yin, Yian},
  journal={Science},
  volume={390},
  number={6779},
  pages={1240--1243},
  year={2025},
  publisher={American Association for the Advancement of Science}
}

@article{hao2026artificial,
  title={Artificial intelligence tools expand scientists’ impact but contract science’s focus},
  author={Hao, Qianyue and Xu, Fengli and Li, Yong and Evans, James},
  journal={Nature},
  pages={1--7},
  year={2026},
  publisher={Nature Publishing Group UK London}
}

@article{gartenberg2026more,
  title={More versus better: Artificial intelligence, incentives, and the emerging crisis in peer review},
  author={Gartenberg, Claudine and Hasan, Sharique and Murray, Alex and Pierce, Lamar},
  journal={Organization Science},
  volume={37},
  number={3},
  pages={795--812},
  year={2026},
  publisher={INFORMS}
}

@article{daniotti2026using,
  title={Who is using AI to code? Global diffusion and impact of generative AI},
  author={Daniotti, Simone and Wachs, Johannes and Feng, Xiangnan and Neffke, Frank},
  journal={Science},
  pages={eadz9311},
  year={2026},
  publisher={American Association for the Advancement of Science}
}

\end{document}